\documentclass[conference,letterpaper]{IEEEtran}

\usepackage{booktabs}
\usepackage{pifont}
\usepackage{graphicx}
\usepackage{amsmath,amssymb}
\usepackage{cite}
\usepackage{url}
\usepackage[hidelinks]{hyperref}
\usepackage{comment}
\usepackage{amsmath,amssymb}
\usepackage[ruled,vlined]{algorithm2e}

\DontPrintSemicolon
\SetAlgoCaptionSeparator{\enspace}
\SetKw{InputLabel}{Input}
\SetKw{OutputLabel}{Output}
\newcommand{\PaperInput}[1]{\InputLabel\enspace #1\;}
\newcommand{\PaperOutput}[1]{\OutputLabel\enspace #1\;}
\SetKwInput{KwIn}{Input}
\SetKwInput{KwOut}{Output}
\SetKw{Return}{return}
\SetKw{Break}{break}

\newcommand{\val}[2]{\ensuremath{#1\pm#2}}
\newcommand{\valb}[2]{\ensuremath{\mathbf{#1}\pm#2}}

\newcommand{\codelink}[1]{\href{#1}{\underline{Link}}}

\title{Beyond Structural Realism: Demographic Homophily Calibration in LLM-Generated Social Networks}

\author{\IEEEauthorblockN{Shikha Mallick\IEEEauthorrefmark{1}, Alex Thomo\IEEEauthorrefmark{1}, and Akrati Saxena\IEEEauthorrefmark{2}}
\IEEEauthorblockA{\IEEEauthorrefmark{1}Department of Computer Science, University of Victoria, BC, Canada\\
\{shikhamallick, thomo\}@uvic.ca}
\IEEEauthorblockA{\IEEEauthorrefmark{2}Leiden Institute of Advanced Computer Science, Leiden University, The Netherlands\\
a.saxena@liacs.leidenuniv.nl}}

\makeatletter
\renewcommand{\@IEEEsectpunct}{.\ \,}
\makeatother

\begin{document}
\maketitle

\begin{abstract}
Large language models (LLMs) offer a flexible way to generate synthetic social networks when empirical network data are unavailable or difficult to access for network science research. However, existing work shows that while LLMs are capable of forming structurally plausible social ties, they misrepresent demographic homophily, potentially distorting simulations of information diffusion and opinion dynamics. To address this limitation, we study calibration toward reference targets for frequently encountered demographics in social networks, such as gender, race/ethnicity, religion, and political affiliation, while aiming to preserve structural integrity. We introduce three calibration methods, namely prompt intervention, LLM fine-tuning using proximal policy optimization (PPO), and target-aware degree-preserving rewiring. We use Llama-3.1-8B-Instruct (Llama) and Gemma-2-9B-IT (Gemma). PPO improves political and race/ethnicity homophily estimates in both models, with larger reductions in both discrepancies for Llama, whose target homophily discrepancies decrease by 87.4\% and 48.8\%, respectively. Applying target-aware rewiring to PPO outputs brings all four mean same-group ratios within 0.03 of their targets in both models. Compared with rewiring unmodified model outputs, rewiring PPO outputs required 32.7\% fewer swaps for Llama and 17.5\% fewer swaps for Gemma while retaining more original ties in both models.
\end{abstract}

\section{Introduction}
Social networks determine the pathways through which people exchange information, encounter different viewpoints, and influence one another~\cite{centola2010spread,dandekar2013biased}. Studying these processes often requires synthetic networks with realistic structural and demographic mixing properties, as empirical data capturing both social ties and demographic attributes is often difficult to obtain due to privacy constraints, access restrictions, and high data collection costs~\cite{chang2025llms,pajola2025elephant,breza2023consistently}. While classical attributed network generative models rely on explicitly parameterized link formation mechanisms~\cite{kim2011modeling,pfeiffer2014attributed}, large language models (LLMs) leverage their pretrained knowledge to generate social networks directly from user descriptions by assuming individual user roles and selecting connections from candidate profiles~\cite{mallick2026llms,chang2025llms,papachristou2025network}. The usefulness of LLM-generated social networks, however, depends on capturing both network topology and the demographic patterns of social interaction. Recent studies show that LLM-generated networks reproduce plausible structural properties while substantially overestimating political homophily, the tendency to connect with others of the same political affiliation~\cite{mallick2026llms,chang2025llms,papachristou2025network}. Such discrepancies alter opportunities for interaction within and across groups, potentially affecting simulations of social influence and information exchange~\cite{mcpherson2001birds,flache2017models}. Mixing patterns also matter for downstream network analysis because homophily can affect how prominently different demographic groups appear in centrality rankings~\cite{saxena2024fairsna}. Empirical homophily varies across attributes and social settings~\cite{mcpherson2001birds,kossinets2009origins,smith2014social,saxena2025homophily}. For example, gender homophily in soccer discussions on Twitter was lower in Portuguese than in English~\cite{macedo2026gender}. Improving demographic fidelity may require either increasing or decreasing attribute-specific homophily in LLM-generated social networks. The objective is therefore agreement with appropriate empirical references rather than uniformly reducing same-group connections.

We investigate how demographic mixing can be calibrated while limiting changes to network structure. To this end, we explore interventions at the input, model, and output stages of LLM-based network generation. These stages offer complementary forms of control, ranging from training-free guidance and reusable policy adaptation to direct graph-level correction. However, demographic calibration may come at the expense of structural fidelity. For instance, although degree-preserving rewiring retains each node's number of connections, it does not necessarily preserve clustering or community structure~\cite{maslov2002specificity}. This motivates a comparative evaluation of demographic calibration and its structural implications across intervention stages.

Our contributions are threefold.
(i) We formulate demographic homophily calibration using population-normalized empirical targets for gender, race/ethnicity, religion, and political affiliation, and introduce a weighted calibration objective, with structural fidelity evaluated separately under a common generation protocol.
(ii) We develop three calibration strategies, namely target-aware prompting, LLM policy adaptation through proximal policy optimization (PPO) with a directional reward that accommodates both increases and decreases in attribute-specific homophily, and target-aware degree-preserving rewiring.
(iii) We empirically evaluate these strategies on a held-out user population using two LLMs, Llama-3.1-8B-Instruct (Llama)~\cite{added_llama3} and Gemma-2-9B-IT (Gemma)~\cite{added_gemma2}. Across both models, PPO improves political and race/ethnicity homophily calibration, while its combination with rewiring brings all four mean homophily ratios within 0.03 of their respective targets. For Llama, PPO reduces political and race/ethnicity discrepancies by 87.4\% and 48.8\%, respectively. PPO also reduced subsequent rewiring by 32.7\% for Llama and 17.5\% for Gemma. These findings demonstrate the potential of learned calibration to reduce reliance on post-generation graph correction.

\section{Related Work}
\subsection{Social Network Generation Using LLMs}
When LLMs select social ties, the resulting networks depend on how the prompt is framed~\cite{mallick2026llms}. For instance, the names assigned to LLM agents can affect how strongly ties concentrate around a few nodes~\cite{demarzo2023emergence}. The type of relationship being modelled also matters, as friendship prompts tend to favour ties between people with a mutual friend, while employment prompts favour well-connected individuals~\cite{papachristou2025network}. Asking the model to select friends for one persona, a user profile with demographic attributes, at a time can yield more realistic network structure than requesting an entire graph in one prompt, yet the resulting friendships can still exhibit substantially more political homophily than observed in empirical networks~\cite{chang2025llms}.

Prompt design also affects demographic groups differently. Providing political or religious information can produce more fragmented networks than providing age or gender, and changing the prompt language can affect religious homophily while leaving political homophily comparatively stable~\cite{mehdizadeh2025homophily,kilaru2026realistic}. These findings show that prompts can influence mixing, but do not tell the model whether a particular friendship choice would move the network closer to an empirical target. Our target-aware prompting method retains sequential friend selection and adds this information.

\subsection{Reinforcement Learning for LLM Fine-Tuning}
Reinforcement learning has been used to adapt language generation to human preferences over continuations and summaries~\cite{ziegler2019finetuning,stiennon2020learning}, as well as to task-specific measures of quality in text-to-graph generation~\cite{dognin2021regen}. In simulated social settings, evaluations of conversational outcomes have also been used to select interactions for subsequent policy training~\cite{wang2024sotopiapi}. The closest precedent for our LLM policy adaptation method is engagement-driven content generation by Coppolillo et al., in which simulated audience responses provide a reward for the posts an LLM produces~\cite{coppolillo2025engagement}.

In engagement-driven generation, the network supplies feedback on content. In our setting, each friendship choice changes the network from which subsequent feedback is computed. A same-group tie can therefore be encouraged while an attribute's mixing ratio is below target and discouraged once it is above target. We use this feedback to fine-tune the LLM's friendship-selection policy with proximal policy optimization (PPO)~\cite{added_ppo}, so the adapted model can apply what it learned without receiving target values in its evaluation prompt.

\subsection{Graph Rewiring}
Degree-preserving edge swaps provide a way to distinguish network patterns explained by the degree sequence from those that depend on how edges are arranged~\cite{maslov2002specificity}. Fast total link rewiring targets a specified degree assortativity by first constructing a highly assortative network with the original degree sequence and then rewiring batches of edges toward the target~\cite{mannion2024fast}. Preserving every node's degree, however, does not fix clustering or path length, as improving agreement on one statistic can change another~\cite{xulvi2004reshuffling,chueca2026structure}. We therefore assess structural preservation separately from demographic calibration.

For demographic calibration, the relevant notion of mixing is similarity in categorical attributes rather than similarity in degree~\cite{newman2003mixing}. Attribute-aware graph generators have used an observed mixing matrix to guide how new subgraphs attach to a growing network~\cite{sikdar2022attributed}. Our rewiring method instead starts from a completed network of persona-selected friendships and chooses degree-preserving swaps that reduce a weighted discrepancy across four demographic targets. Applying the same correction to graphs from different generation methods lets us distinguish outcomes that final mixing ratios alone would obscure, as two methods may reach comparable demographic agreement while requiring different numbers of swaps and retaining different fractions of their original friendships.

\section{Problem Definition}
We calibrate demographic mixing toward reference targets while limiting structural distortion in LLM-generated social networks.

\subsection{Attributed Social Networks}
In this work, we define a social network as a simple, undirected attributed graph $G=(V,E,X)$, where the node set $V=\{1,\ldots,n\}$ represents $n$ user profiles, or personas, each described by demographic attributes. The edge set $E\subseteq\{\{u,v\}\mid u,v\in V,u\neq v\}$ represents friendships. Let $\mathcal{A}$ be the categorical demographic attributes, with categories $\mathcal{C}_a$ for each attribute $a$. The attribute table is $X=[x_v^{(a)}]_{v\in V,a\in\mathcal{A}}$, where $x_v^{(a)}\in\mathcal{C}_a$ denotes the category of attribute $a$ assigned to persona $v$. We calibrate $\mathcal{A}_{\mathrm{cal}}\subseteq\mathcal{A}$ comprising gender, race/ethnicity, religion, and political affiliation. Let $m=|E|$ and assume $n\geq2$. Edge fractions below require $m>0$.

\subsection{Structural Metrics}
\label{sm}
To evaluate whether generated networks satisfy realistic structural properties, we measure density, average local clustering~\cite{watts1998smallworld}, modularity of the community partition identified by the Louvain algorithm with resolution parameter $\gamma$~\cite{newman2004community,added_louvain}, the fraction of nodes in the largest connected component (LCC), and average shortest-path length (APL), and compare against real references. Local clustering is zero for nodes of degree below two. APL is computed on the largest connected component and normalized by the natural logarithm of its size, with singleton values set to zero.

For clarity, let $k_v$ be node $v$'s degree, $t_v$ the number of edges among its neighbours, and $V_{\max}$ the nodes in a largest connected component. Let $c_v=2t_v/[k_v(k_v-1)]$ for $k_v\geq2$ and $c_v=0$ otherwise. Density $D(G)$ and average local clustering $C(G)$ are
$$
D(G)=\frac{2m}{n(n-1)},\qquad
C(G)=\frac{1}{n}\sum_{v\in V}c_v.
$$
Connectivity is $L(G)=|V_{\max}|/n$. If $s=|V_{\max}|>1$, let $d_G(u,v)$ be the shortest-path distance between $u$ and $v$. The normalized average path length is
$$
A(G)=\frac{1}{s(s-1)\log s}
\sum_{u\in V_{\max}}\sum_{\substack{v\in V_{\max}\\v\neq u}}d_G(u,v).
$$
Modularity evaluates the Louvain partition through excess within-community connectivity relative to the degree-based null model~\cite{newman2004community,added_louvain}. These summaries capture distinct properties, and none alone establishes that a generated graph is realistic.

\subsection{Demographic Homophily}
\label{dm}
Generated networks may be structurally realistic, but for attributed networks, the demographic mixing patterns must be realistic as well. In this work, we evaluate demographic mixing using the same-group edge fraction. Let $\mathbf{1}[\cdot]$ denote the indicator function. For $a\in\mathcal{A}_{\mathrm{cal}}$, the observed same-group edge fraction is
\begin{equation}
\label{eq:sgr}
h_a(G)=\frac{1}{m}\sum_{\{u,v\}\in E}\mathbf{1}[x_u^{(a)}=x_v^{(a)}].
\end{equation}
To account for population composition, let $n_{a,c}$ count nodes in category $c$ and define
\begin{equation}
\label{eq:same-cat}
q_a=\frac{\sum_{c\in\mathcal{C}_a}n_{a,c}(n_{a,c}-1)}{n(n-1)},
\end{equation}
the probability that two distinct uniformly sampled nodes share a category. For $0<q_a<1$, the normalized same-group ratio (SGR) is
\begin{equation}
\label{eq:norm-sgr}
H_a^{\mathrm{same}}(G)=h_a(G)/q_a.
\end{equation}
Values higher or lower than one indicate over-representation or under-representation of same-group ties relative to this baseline. We calibrate against empirical reference targets specified in Chang et al.~\cite{chang2025llms} (Section~\ref{exp}), which need not equal one, meaning that lower homophily is not necessarily more realistic. For example, an SGR of 1.4 corresponds to 70\% within-group edges when half of all candidate pairs share a category, but only 35\% when one quarter do.

\subsection{LLM-Based Network Generation}
\label{sec:generation}
Given a population of personas $V$ and their demographic attributes $X$, we construct a social network $G$ using an LLM. Starting from an empty graph, we process personas sequentially. At each step, the model is prompted to impersonate the focal persona and receives that persona's profile, eligible candidate profiles, and a requested number of friend selections. An abbreviated LLM prompt reads
\emph{``You are [persona profile]. Select [requested number] friends from [candidate profiles and available network context]. Return the selected candidate identifiers.''} Eligible candidates exclude the focal persona and its existing neighbours. During an initial phase, focal personas select from profile information alone, whereas subsequent prompts also include each candidate's current degree in the evolving graph. Valid selections are added as undirected edges, so a friendship exists if either persona selects the other. The requested number of selections need not equal a persona's final degree, since other personas may also select them. Generation ends after processing all personas. We denote the unmodified instruction-tuned model by \textbf{Base} and a graph it generates by $G_{\mathrm{base}}$. Across methods, personas, attributes, and generation protocol remain fixed except for intervention-specific changes. Prompting changes the friend-selection instructions, PPO adapts the model parameters, and target-aware degree-preserving rewiring modifies the completed graph.

In Algorithm~\ref{alg:base}, $\sigma_t$ is the persona at step $t$, $b_i$ is persona $i$'s selection budget, and $t_0$ is the number of initial steps without degree information. For focal persona $i$, let $N_G(i)$ denote its neighbours, $C$ its eligible candidates, $x_i$ its profile, and $X_C$ the candidate profiles. The observation $o$ includes the available degree information $D$, where $\deg_G(j)$ is candidate $j$'s degree. The policy $\pi$ samples a response $y$ to the prompt constructed by $\mathcal{P}$. The procedure $\textsc{ValidIDs}$ retains the distinct eligible identifiers in $y$ as the set $A$. We write $\textsc{Generate}(\pi,\mathcal{P})$ for this algorithm with the specified policy and prompt constructor.

\begin{algorithm}[!t]
\caption{Base network generation}
\label{alg:base}

\PaperInput{Personas $(V,X)$, order $\sigma$, budgets $(b_i)_{i\in V}$,
      policy $\pi$, prompt constructor $\mathcal{P}$}
\PaperOutput{Undirected graph $G=(V,E,X)$}

$E \gets \varnothing$\;
\For{$t \gets 1$ \KwTo $|V|$}{
    $i \gets \sigma_t$, \quad $G \gets (V,E,X)$\;
    $C \gets V \setminus \bigl(\{i\}\cup N_G(i)\bigr)$\;
    $D \gets \varnothing$\;
    \If{$t>t_0$}{
        $D \gets \{(j,\deg_G(j))\mid j\in C\}$\;
    }
    $o \gets (x_i,X_C,b_i,D)$\;
    $y \sim \pi\bigl(\cdot\mid\mathcal{P}(o,G)\bigr)$\;
    $A \gets \textsc{ValidIDs}(y,C)$\;
    $E \gets E \cup \{\{i,j\}\mid j\in A\}$\;
}
\Return $(V,E,X)$\;
\end{algorithm}

\begin{figure*}[t]
\centering\includegraphics[width=\linewidth]{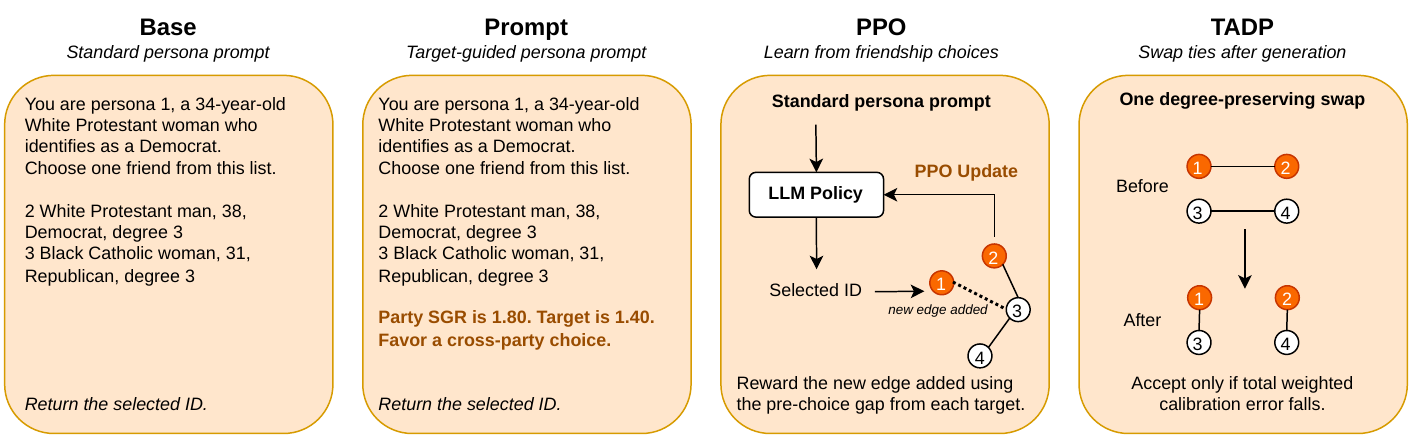}
\caption{The original network generative method ``Base" and our three intervention methods, ``Prompt", ``PPO", and ``TADP", illustrated side by side. Base and Prompt show abbreviated instructions for the same persona and candidates. Prompt adds demographic guidance, illustrated here for political affiliation. During PPO training, the model selects a friend and receives a reward based on how the new tie aligns with the demographic targets given the graph before selection. The dashed edge shows the selected tie. TADP instead edits a completed graph by swapping two ties when doing so reduces total calibration error while preserving every node's degree. Filled and open nodes denote different political affiliations.}
\label{fig:paths}\end{figure*}

\subsection{Calibration Objective}
\label{calib-obj}
For each attribute $a$, let $\tau_a$ be its reference SGR target and $w_a$ its fixed priority weight. We define demographic calibration error as the weighted absolute deviation of SGRs from their targets as follows
$$
B(G)=\sum_{a\in\mathcal{A}_{\mathrm{cal}}}
w_a\left|H_a^{\mathrm{same}}(G)-\tau_a\right|.
$$
Let $\bar H_a$ be the corresponding model's mean Base SGR on the training population and $\tau_{a,j}$ the reference for dataset/relationship unit $j$, with $J_a$ units. Let $d_a$ denote the mean absolute discrepancy across these units and let $b$ index calibrated attributes in the normalization. The effective fixed weights are
$$
d_a=\frac{1}{J_a}\sum_{j=1}^{J_a}|\bar H_a-\tau_{a,j}|,
\qquad w_a=\frac{d_a}{\sum_{b\in\mathcal{A}_{\mathrm{cal}}}d_b}.
$$
The weights are fixed using the training population and remain unchanged during training and evaluation. Target-aware rewiring directly reduces this error, whereas PPO uses a directional edge-based reward. Minimizing $B(G)$ does not imply that every attribute improves at every step. A reduction in one attribute’s weighted absolute deviation from its target can offset an increase in another attribute.

\section{Demographic Calibration Methods}
We investigate three approaches to demographic calibration, namely prompt intervention before generation, LLM policy adaptation through fine-tuning, and graph correction after generation.

\subsection{Prompt Intervention}
We augment the friendship-selection prompt with target and current homophily ratios for all four attributes, together with their priority weights. The instructions encourage cross-group choices when homophily exceeds its target and same-group choices when it falls below. This provides target-aware guidance without updating model parameters. For example, the prompt adds
\emph{``Current same-group ratios are [ratios]. Reference targets are [targets]. Attribute priorities are [weights]. Favor cross-group choices for attributes above their targets and same-group choices for attributes below their targets.''}

In Algorithm~\ref{alg:prompt}, $\pi_0$ and $\mathcal{P}_0$ are the frozen Base policy and standard prompt constructor. The guided constructor $\mathcal{P}_{\mathrm{guide}}$ builds prompt text $p$. For each attribute, $H$ is its current SGR or $\bot$ if undefined. The instruction $d$ is empty unless it favors cross-group or same-group ties. The operator $\Vert$ appends the text produced by $\textsc{Format}$.

\begin{algorithm}[!t]
\caption{Target-guided prompt generation}
\label{alg:prompt}

\PaperInput{Frozen policy $\pi_0$, standard prompt $\mathcal{P}_0$,
      targets $(\tau_a)$, weights $(w_a)$}
\PaperOutput{Prompt-generated graph $G$}

\SetKwProg{Fn}{Function}{}{}
\Fn{$\mathcal{P}_{\mathrm{guide}}(o,G)$}{
    $p \gets \mathcal{P}_0(o,G)$\;
    \ForEach{$a\in\mathcal{A}_{\mathrm{cal}}$}{
        $H \gets H_a^{\mathrm{same}}(G)$\;
        $d \gets \varnothing$\;
        \If{$H\neq\bot$}{
            \uIf{$H>\tau_a$}{
                $d \gets \textsc{FavorCross}$\;
            }
            \ElseIf{$H<\tau_a$}{
                $d \gets \textsc{FavorSame}$\;
            }
        }
        $p \gets p \mathbin{\Vert}
            \textsc{Format}(a,H,\tau_a,w_a,d)$\;
    }
    \Return $p$\;
}
\BlankLine
\Return $\textsc{Generate}(\pi_0,\mathcal{P}_{\mathrm{guide}})$\;
\end{algorithm}

\subsection{PPO Fine-Tuning}
We adapt the LLM using proximal policy optimization (PPO)~\cite{added_ppo}. PPO retains the friend-selection prompt from Section~\ref{sec:generation} and learns calibration preferences through parameter updates. During sequential network generation, the PPO reward favors friendship choices toward each demographic target, emphasizing attributes with larger discrepancies and higher weights. Let $G_{t-1}$ denote the graph before generation step $t$, and let $e_{a,t}=H_a^{\mathrm{same}}(G_{t-1})-\tau_a$ denote the discrepancy for attribute $a$ before step $t$. Let $k_t$ be the number of newly added edges. The coefficients $\alpha,\beta>0$ control the reward and penalty magnitudes. When $e_{a,t}>0$, $T_{a,t}$ counts new cross-group edges and $U_{a,t}$ counts new same-group edges. When $e_{a,t}<0$, these roles are reversed. Attributes with zero or undefined discrepancy contribute zero. We define the step-level demographic reward $r_t$ as
\begin{equation}\label{eq:ppo-reward}
r_t=
\frac{1}{\max(k_t,1)}
\sum_{a\in\mathcal{A}_{\mathrm{cal}}}
w_a |e_{a,t}|
\left(\alpha T_{a,t}-\beta U_{a,t}\right).
\end{equation}
The reward supports both increases and decreases in homophily, encouraging target agreement without directly minimizing the graph-level error $B(G)$. We fine-tune LoRA adapters~\cite{hu2022lora} with a value head and a KL penalty toward the frozen Base policy~\cite{ziegler2019finetuning}. We penalize unparseable responses, sample training rollouts, and select the checkpoint by calibration error on validation seeds disjoint from the evaluation seeds.

In Algorithm~\ref{alg:ppo}, $\phi$ parameterizes the adapters of policy $\pi_\phi$, and $\psi$ parameterizes its value head $v_\psi$. The rollout procedure samples a batch $\mathcal{D}$ of responses $y_t$ on training population $V_{\mathrm{tr}}$. An unparseable response receives the failure reward $r_{\mathrm{fail}}$. The PPO update uses the batch rewards and frozen reference policy. Among saved checkpoints $\phi'\in\mathcal{K}$, the selected checkpoint $\phi^\star$ minimizes mean calibration error $\widehat{B}_{\mathrm{val}}$ over validation seeds $\mathcal{S}_{\mathrm{val}}$. The reward scores the direction of a selection using the graph before the choice. A multi-edge choice can cross a target or improve one attribute at the expense of another. PPO therefore need not reduce $B(G)$ at every step.

\begin{algorithm}[!t]
\caption{PPO policy adaptation and network generation}
\label{alg:ppo}

\PaperInput{Frozen policy $\pi_0$, training population $V_{\mathrm{tr}}$,
      validation seeds $\mathcal{S}_{\mathrm{val}}$,
      failure reward $r_{\mathrm{fail}}<0$}
\PaperOutput{Graph generated by the selected PPO policy}

Initialize LoRA parameters $\phi$ and value head $v_\psi$\;
$\mathcal{K}\gets\varnothing$\;

\ForEach{training batch}{
    $\mathcal{D}\gets
      \textsc{Rollout}(V_{\mathrm{tr}},\pi_\phi,\mathcal{P}_0)$\;
    \ForEach{response $y_t$ in $\mathcal{D}$}{
        \eIf{$y_t$ is parseable}{
            Compute $r_t$ from $G_{t-1}$ and the new ties
            using \eqref{eq:ppo-reward}\;
        }{
            $r_t\gets r_{\mathrm{fail}}$\;
        }
        Store $r_t$ with $y_t$ in $\mathcal{D}$\;
    }
    $(\phi,\psi)\gets
      \textsc{PPOUpdate}(\phi,\psi,\mathcal{D},\pi_0)$\;
    \If{a checkpoint is due}{
        $\mathcal{K}\gets\mathcal{K}\cup\{\phi\}$\;
    }
}
$\phi^\star\in\arg\min_{\phi'\in\mathcal{K}}
  \widehat{B}_{\mathrm{val}}(\phi',\mathcal{S}_{\mathrm{val}})$\;
\Return $\textsc{Generate}(\pi_{\phi^\star},\mathcal{P}_0)$\;
\end{algorithm}

\subsection{Target-Aware Degree-Preserving Rewiring}
Target-aware degree-preserving rewiring (TADP) refines graphs generated by Base, Prompt, or PPO through degree-preserving edge swaps~\cite{maslov2002specificity}. At each step, TADP evaluates a bounded number of sampled candidate swaps, applies the one that most reduces the weighted total calibration error $B(G)$ across all four attributes (individual attribute errors may increase if the total decreases), and stops when no sampled swap improves $B(G)$ or the operation budget is reached. Each accepted swap replaces two edges, keeps the graph simple, and preserves every node's degree, the edge count, and density exactly. For four distinct endpoints $u,v,x,y\in V$, replacing $\{u,v\}$ and $\{x,y\}$ by $\{u,x\}$ and $\{v,y\}$ preserves all four endpoint degrees when the new edges are valid and the endpoints are distinct. Self-loops, duplicate edges, and invalid swaps are rejected. With fixed node attributes and edge count, only the removed and added edges affect the same-group counts. Let $s_a$ be the number of same-group edges for attribute $a$ and let $\Delta s_a$ be its change under a candidate swap. The resulting change in SGR is
$$
\Delta H_a^{\mathrm{same}}=\frac{\Delta s_a}{m q_a}.
$$
A candidate's objective can therefore be evaluated from its changed edges without recomputing every homophily statistic from scratch. The implementation maintains these counts while assessing sampled moves. Stopping after no sampled improvement means that the sampled candidates did not improve the objective. It is not a certificate of global optimality or even an exhaustive local optimum. Different random candidates may yield different endpoints, and fixed degrees can prevent exact realization of arbitrary targets. The shared operation budget makes the amount of permitted intervention explicit while actual accepted operations are measured separately.

In Algorithm~\ref{alg:tadp}, $k$ indexes swap attempts, $K$ bounds accepted swaps, and $\mathcal{S}$ contains up to $S$ sampled candidates. For a valid swap $s$, $G\oplus s$ denotes the resulting graph and $\Delta_s$ the decrease in $B$. The algorithm selects a candidate $s^\star$ with the largest decrease.

\begin{algorithm}[!t]
\caption{Target-aware degree-preserving rewiring}
\label{alg:tadp}

\PaperInput{Initial graph $G$, calibration error $B$,
      candidate limit $S$, accepted-swap budget $K$}
\PaperOutput{Corrected graph with the original degree sequence}

\For{$k\gets 1$ \KwTo $K$}{
    $\mathcal{S}\gets\textsc{SampleSwaps}(G,S)$\;
    $\mathcal{S}\gets
      \{s\in\mathcal{S}\mid \textsc{ValidSwap}(G,s)\}$\;
    \If{$\mathcal{S}=\varnothing$}{
        \Break\;
    }
    \ForEach{$s\in\mathcal{S}$}{
        $\Delta_s\gets B(G)-B(G\oplus s)$\;
    }
    $s^\star\in\arg\max_{s\in\mathcal{S}}\Delta_s$\;
    \If{$\Delta_{s^\star}\leq 0$}{
        \Break\;
    }
    $G\gets G\oplus s^\star$\;
}
\Return $G$\;
\end{algorithm}

Random degree-preserving swaps isolate the effects of maintaining degrees without using demographic targets. Target-aware edge replacement isolates target guidance without the degree constraint. TADP combines these two choices. Applying it to a PPO-generated graph leaves the graph-correction algorithm unchanged, allowing us to examine whether learned calibration supplies an input that needs fewer corrections. A reduction in swaps measures post-generation editing effort, not the total computational cost of training and inference. Figure~\ref{fig:paths} illustrates friendship selection during generation and edge swaps after generation.

\begin{table*}[t]
\centering
\caption{Llama-3.1-8B on the 50-person held-out evaluation population with structural metrics (top) and same-group ratios (bottom), mean plus or minus one sample standard deviation (SD) over 30 graphs. Bold marks the mean(s) closest to the corresponding reference at the displayed precision, including all ties.}
\label{tab:llama_50}
\begin{tabular*}{\textwidth}{@{\extracolsep{\fill}}lccccccc@{}}
\toprule
 & & \multicolumn{2}{c}{Base} & \multicolumn{2}{c}{Prompt-guided} & \multicolumn{2}{c}{PPO-adapted} \\
\cmidrule(lr){3-4}\cmidrule(lr){5-6}\cmidrule(lr){7-8}
Metric & Reference & Base & Base-TADP & Prompt & Prompt-TADP & PPO & PPO-TADP \\
\midrule
Density & 0.19 & \valb{0.15}{0.02} & \valb{0.15}{0.02} & \valb{0.15}{0.02} & \valb{0.15}{0.02} & \valb{0.15}{0.02} & \valb{0.15}{0.02} \\
Clustering & 0.45 & \valb{0.44}{0.05} & \val{0.43}{0.06} & \val{0.38}{0.06} & \val{0.47}{0.07} & \val{0.34}{0.06} & \val{0.43}{0.09} \\
Modularity & 0.39 & \valb{0.38}{0.04} & \val{0.35}{0.03} & \val{0.31}{0.03} & \val{0.36}{0.03} & \val{0.30}{0.03} & \val{0.35}{0.04} \\
LCC & 0.99 & \valb{1.00}{0.00} & \valb{1.00}{0.00} & \valb{1.00}{0.00} & \valb{1.00}{0.01} & \valb{1.00}{0.01} & \valb{1.00}{0.01} \\
Norm. APL & 0.65 & \valb{0.59}{0.03} & \val{0.58}{0.03} & \val{0.56}{0.03} & \val{0.57}{0.03} & \val{0.56}{0.03} & \val{0.57}{0.03} \\
\midrule
Gender & 1.07 & \val{1.11}{0.07} & \valb{1.07}{0.01} & \val{1.08}{0.04} & \valb{1.07}{0.01} & \val{1.04}{0.06} & \valb{1.07}{0.01} \\
Race/ethnicity & 1.68 & \val{1.22}{0.12} & \val{1.65}{0.07} & \val{1.29}{0.17} & \val{1.62}{0.10} & \val{1.44}{0.12} & \valb{1.67}{0.02} \\
Religion & 1.77 & \val{1.60}{0.11} & \val{1.74}{0.08} & \val{1.48}{0.15} & \val{1.74}{0.09} & \val{1.48}{0.12} & \valb{1.77}{0.02} \\
Political affil. & 1.40 & \val{2.23}{0.16} & \valb{1.40}{0.00} & \val{1.70}{0.14} & \valb{1.40}{0.00} & \val{1.51}{0.13} & \valb{1.40}{0.00} \\
\bottomrule
\end{tabular*}
\end{table*}

\begin{table*}[t]
\centering
\caption{Gemma-2-9B on the 50-person held-out evaluation population with structural metrics (top) and same-group ratios (bottom), mean plus or minus one sample standard deviation (SD) over 30 graphs. Bold marks the mean(s) closest to the corresponding reference at the displayed precision, including all ties.}
\label{tab:gemma9b_50}
\begin{tabular*}{\textwidth}{@{\extracolsep{\fill}}lccccccc@{}}
\toprule
 & & \multicolumn{2}{c}{Base} & \multicolumn{2}{c}{Prompt-guided} & \multicolumn{2}{c}{PPO-adapted} \\
\cmidrule(lr){3-4}\cmidrule(lr){5-6}\cmidrule(lr){7-8}
Metric & Reference & Base & Base-TADP & Prompt & Prompt-TADP & PPO & PPO-TADP \\
\midrule
Density & 0.19 & \valb{0.15}{0.02} & \valb{0.15}{0.02} & \val{0.14}{0.02} & \val{0.14}{0.02} & \valb{0.15}{0.02} & \valb{0.15}{0.02} \\
Clustering & 0.45 & \val{0.48}{0.06} & \val{0.44}{0.08} & \val{0.38}{0.07} & \val{0.44}{0.07} & \valb{0.45}{0.07} & \valb{0.45}{0.08} \\
Modularity & 0.39 & \val{0.40}{0.04} & \val{0.36}{0.04} & \val{0.36}{0.03} & \valb{0.39}{0.03} & \val{0.36}{0.04} & \val{0.37}{0.03} \\
LCC & 0.99 & \valb{1.00}{0.00} & \valb{1.00}{0.00} & \valb{1.00}{0.00} & \valb{1.00}{0.00} & \valb{1.00}{0.00} & \valb{1.00}{0.01} \\
Norm. APL & 0.65 & \valb{0.61}{0.04} & \val{0.59}{0.03} & \val{0.59}{0.03} & \val{0.59}{0.03} & \val{0.60}{0.03} & \val{0.59}{0.03} \\
\midrule
Gender & 1.07 & \val{1.14}{0.06} & \valb{1.07}{0.01} & \val{1.19}{0.05} & \valb{1.07}{0.01} & \val{1.12}{0.06} & \valb{1.07}{0.01} \\
Race/ethnicity & 1.68 & \val{1.25}{0.12} & \val{1.64}{0.08} & \val{1.29}{0.14} & \val{1.64}{0.06} & \val{1.37}{0.12} & \valb{1.66}{0.03} \\
Religion & 1.77 & \val{1.71}{0.13} & \valb{1.75}{0.05} & \val{1.59}{0.13} & \val{1.74}{0.06} & \val{1.69}{0.13} & \valb{1.75}{0.04} \\
Political affil. & 1.40 & \val{2.22}{0.14} & \valb{1.40}{0.00} & \val{2.01}{0.12} & \valb{1.40}{0.01} & \val{1.98}{0.14} & \valb{1.40}{0.01} \\
\bottomrule
\end{tabular*}
\end{table*}

\section{Experiments}
\label{exp}
\subsection{Experimental Setup}
\paragraph{Personas and generation}
We construct two 50-persona populations, one for training and one excluded from PPO training (held-out), by combining 2023 U.S. Census demographic information~\cite{added_census} with respondent records from the 2024 Cooperative Election Study (CES)~\cite{added_ces2024}. Religion and political affiliation are sampled jointly within demographic groups, preserving their survey association. Profiles include age, gender (Census binary sex), race/ethnicity, religion, and political affiliation, but names and interests are excluded. PPO uses the training population, and all methods are evaluated on the separately sampled held-out population. We generate 30 graphs per condition using the sequential procedure in Section~\ref{sec:generation}. The first $t_0=3$ personas select from profiles alone. Requested friend counts follow a clipped exponential distribution with scale 4 and range 1-15. Inference temperatures, 0.1 for Llama and 1.0 for Gemma, were selected per model on separate selection seeds before the held-out runs, while other generation settings and output validation are shared across conditions.

\paragraph{Models and conditions}
For each model, we compare Base with \textbf{Prompt} (prompt intervention), which adds target guidance, and \textbf{PPO}, which uses the adapted policy. Applying TADP to each generator gives \textbf{Base-TADP}, \textbf{Prompt-TADP}, and \textbf{PPO-TADP}, respectively. Controls include \textbf{RR}, which randomly replaces edges, \textbf{DP}, which randomly swaps edges while preserving node degrees, and \textbf{TARR}, which replaces edges to reduce calibration error without preserving node degrees. We use LoRA rank 16, scaling 32, and dropout 0.05. The PPO reward uses $\alpha=5$, $\beta=0.5$, and $r_{\mathrm{fail}}=-0.5$ with up to four retries. Training rollouts use temperature 1.0. TADP samples up to $S=5{,}000$ candidates per iteration. Rewiring uses a budget of accepted operations equal to approximately 20\% of the input edge count.

\paragraph{Evaluation}
We derive demographic reference targets from the empirical homophily benchmarks compiled by Chang et al.~\cite{chang2025llms}. We fix the same-group ratio (SGR) targets at 1.0687 for gender, 1.6769 for race/ethnicity, 1.7733 for religion, and 1.4036 for political affiliation. Gender references come from General Social Survey (GSS) discussion networks~\cite{smith2014social}, MySpace friendships~\cite{thelwall2009homophily}, and Tuenti friendship and interaction networks~\cite{laniado2016gender}. Race/ethnicity and religion references come from GSS and MySpace. Political affiliation references come from Twitter following networks~\cite{halberstam2016homophily}. Each target averages SGRs within dataset-relationship combinations, then equally across combinations. Thus, datasets with more network samples do not receive greater weight. The targets describe mixing in these source networks rather than national demographic proportions. For structural evaluation, we use the empirical network collection described by Chang et al.~\cite{chang2025llms}, selecting graphs with between half and twice the number of nodes in our generated networks. Structural references use the same averaging procedure, weighting combinations of dataset and relationship type equally. Structural and demographic references therefore come from separately selected collections of empirical networks. We report means and sample standard deviations (SD) across 30 graphs. Tables~\ref{tab:llama_50} and~\ref{tab:gemma9b_50} bold all means closest to the displayed row reference. SD does not break ties. We distinguish structural preservation relative to the input graph from agreement with empirical references.

\paragraph{Reproducibility and scope}
The released configurations identify the selected Llama and Gemma adapters at episodes 39 and 69, respectively. Held-out graphs use 30 generation seeds beginning at 6000, and training-population controls use seeds beginning at 1000. The release includes persona files, graph adjacency lists, per-graph measurements, summary tables, and a CPU-only verification script. Structural calculations use NetworkX~\cite{hagberg2008networkx}. Louvain uses resolution $\gamma=1$ and a fixed random seed. These artifacts allow verification of reported graph statistics without rerunning language-model training.

\subsection{Results}
In Tables~\ref{tab:llama_50} and~\ref{tab:gemma9b_50}, structural rows follow Section~\ref{sm} (\textbf{Norm. APL} is normalized average shortest-path length), demographic rows report SGRs, and \textbf{Reference} is the empirical reference value.

\paragraph{Demographic calibration}
Before rewiring, both Prompt and PPO move political and race/ethnicity SGRs toward their targets, with PPO closer on both attributes in both models (Tables~\ref{tab:llama_50} and~\ref{tab:gemma9b_50}). For Llama, PPO shifts political SGR from 2.23 to 1.51 (target 1.40) and race/ethnicity SGR from 1.22 to 1.44 (target 1.68). The corresponding Gemma changes are 2.22 to 1.98 and 1.25 to 1.37. Gender favors Prompt over PPO for Llama and PPO over Prompt for Gemma. Llama PPO moves religion farther from its 1.77 target (1.60 to 1.48), possibly because religion and party affiliation are sampled jointly. After rewiring, all four PPO-TADP mean SGRs lie within 0.03 of their targets in both models. All three rewired conditions match the displayed gender and political targets in both models. PPO-TADP is closest for race/ethnicity in both. Prompt-TADP reaches 1.62 for Llama and 1.64 for Gemma, compared with 1.67 and 1.66 for PPO-TADP. For religion, PPO-TADP is closest for Llama and ties Base-TADP at 1.75 for Gemma.

\begin{table*}[t]
\centering
\caption{Llama rewiring controls, means over 30 graphs per condition. For each metric, bold marks the method mean closest to the empirical reference, evaluated separately for the training and held-out datasets. All ties at the displayed precision are bolded. Each corrected graph starts from the corresponding Base graph. Race abbreviates race/ethnicity and politics denotes political affiliation.}
\label{tab:ablation}
\begin{tabular*}{\textwidth}{@{\extracolsep{\fill}}lccccccc@{}}
\toprule Dataset & Method & Gender & Race & Religion & Politics & Clustering & Modularity \\
\midrule
Empirical reference & - & 1.07 & 1.68 & 1.77 & 1.40 & 0.45 & 0.39 \\
\midrule
Training & Base & 1.10 & 1.24 & 1.83 & 2.14 & 0.43 & \textbf{0.37} \\
Training & RR & 1.06 & 1.19 & 1.66 & 1.89 & 0.30 & 0.32 \\
Training & DP & \textbf{1.07} & 1.13 & 1.46 & 1.67 & 0.30 & 0.26 \\
Training & TARR & \textbf{1.07} & \textbf{1.61} & \textbf{1.77} & 1.53 & 0.31 & 0.31 \\
Training & TADP & \textbf{1.07} & 1.59 & \textbf{1.77} & \textbf{1.40} & \textbf{0.44} & \textbf{0.37} \\
\midrule
Held-out & Base & 1.11 & 1.22 & 1.60 & 2.23 & \textbf{0.44} & \textbf{0.38} \\
Held-out & RR & 1.08 & 1.18 & 1.47 & 1.96 & 0.31 & 0.33 \\
Held-out & DP & 1.08 & 1.15 & 1.35 & 1.70 & 0.30 & 0.27 \\
Held-out & TARR & \textbf{1.07} & 1.63 & \textbf{1.78} & 1.62 & 0.32 & 0.31 \\
Held-out & TADP & \textbf{1.07} & \textbf{1.65} & 1.74 & \textbf{1.40} & 0.43 & 0.35 \\
\bottomrule
\end{tabular*}
\end{table*}
\begin{table}[t]
\centering
\caption{Editing effort on the held-out population (30 graphs per condition). Within each model, bold marks fewer swaps, fewer original edges removed, or more retained. Edge percentages are computed relative to each input graph.}
\label{tab:effort}
\begin{tabular*}{\linewidth}{@{\extracolsep{\fill}}llrr@{}}
\toprule
Model & Metric & Base-TADP & PPO-TADP \\
\midrule
Llama & Accepted swaps
      & 32.27 & \textbf{21.73} \\
      & Original edges removed
      & 32.45\% & \textbf{20.99\%} \\
      & Original edges retained
      & 67.55\% & \textbf{79.01\%} \\
\midrule
Gemma & Accepted swaps
      & 30.07 & \textbf{24.80} \\
      & Original edges removed
      & 30.71\% & \textbf{24.89\%} \\
      & Original edges retained
      & 69.29\% & \textbf{75.11\%} \\
\bottomrule
\end{tabular*}
\end{table}

\paragraph{Differences across generators}
The unmodified models have almost identical political SGRs (2.23 and 2.22), yet Llama responds more strongly to both interventions. A larger initial political discrepancy therefore does not explain the difference. The model comparison also includes different decoding temperatures and does not isolate architecture.

\paragraph{PPO reduces subsequent rewiring}
Table~\ref{tab:effort} compares the editing effort required when TADP starts from Base or PPO outputs. Starting from PPO outputs reduced the mean number of accepted swaps per graph from 32.27 to 21.73 for Llama and from 30.07 to 24.80 for Gemma, reductions of 32.7\% and 17.5\%, respectively. The fraction of original ties retained increased from 67.55\% to 79.01\% for Llama and from 69.29\% to 75.11\% for Gemma. Retention complements the swap count because a later swap can alter a previously edited tie.

\paragraph{Structural preservation}
The structural effects depend on the model (Tables~\ref{tab:llama_50} and~\ref{tab:gemma9b_50}). For Llama, PPO-TADP increases clustering from PPO's 0.34 to 0.43 and modularity from 0.30 to 0.35, matching Base-TADP at the reported precision. For Gemma, PPO already attains clustering of 0.45, matching the empirical reference at the reported precision. Rewiring retains this value and increases modularity from 0.36 to 0.37. Prompt-TADP raises clustering from 0.38 to 0.47 for Llama and to 0.44 for Gemma. Base remains closest to the normalized APL reference in both models and to the modularity reference for Llama. For Gemma, Prompt-TADP matches the displayed modularity reference of 0.39. TADP preserves input degrees and density, and all six conditions have mean LCC fractions of 1.00 at the displayed precision. Degree preservation restricts which discrepancies rewiring can correct. For instance, it can change triangles, but it cannot compensate for an initial edge budget that yields a density below the reference.

\paragraph{Rewiring ablation}
We compared RR, DP, TARR, and TADP for Llama under the same budget (Table~\ref{tab:ablation}). On the held-out population, TADP was closest to the race/ethnicity and political targets and tied TARR on gender. TARR was closer on religion, reaching 1.78 against a target of 1.77, compared with 1.74 for TADP. Among the rewiring methods, TADP also came closest to the empirical clustering and modularity references, reaching 0.43 and 0.35, respectively. On the training population, TADP retained higher clustering than TARR (0.44 versus 0.31) and better matched political homophily (1.40 versus 1.53), although TARR better matched race/ethnicity. This balance motivates TADP as our graph-correction method.

\paragraph{Comparison with an independent-edge model}
An independent-edge logistic model fitted to reproduce the four SGR targets and the reference density matches the targets on the training population (mean absolute SGR discrepancy below 0.01) but reaches clustering of only 0.25 and modularity of 0.26 against references of 0.45 and 0.39, even with a latent four-block community offset. Demographic summaries alone thus do not yield realistic structure.

\section{Discussion}
The choice of calibration method depends on whether the goal is to guide friendship choices during generation or to correct a completed network. Prompting provides target guidance without training, while PPO adapts the selection policy for reuse across generation runs. When the generated graph can be edited, target-aware degree-preserving rewiring (TADP) brings demographic mixing close to the empirical targets while preserving each persona's degree. For both models, starting TADP from PPO outputs required fewer swaps and preserved more original friendships, with a larger gain in retention for Llama. The final mixing ratios alone obscure this benefit of policy adaptation.

The independent-edge model closely matched the demographic targets but produced substantially less clustering than the empirical reference. Networks with similar rates of same-group friendship can therefore differ in how often two friends of a persona are also friends with each other. For a diffusion simulation, this distinction matters because mutual friendships can provide multiple paths through which the same information reaches a persona. Evaluating both mixing and clustering helps distinguish generators that appear equally suitable on demographic grounds.

The empirical targets describe mixing in the populations and relationships represented in the source datasets, including friendship, discussion, and following. They should not be interpreted as universal benchmarks. The 30 runs per condition capture variation across generations on the same 50 held-out personas. Repeating the comparison on independently sampled populations would show how sensitive the findings are to population composition. Experiments on diffusion and opinion dynamics could test when closer demographic agreement changes the conclusions drawn from a simulation.

\section{Conclusion}
LLM-generated networks exhibit structural realism yet misrepresent who connects across demographic groups. We examined how to calibrate those opportunities by changing the instructions given to an LLM, adapting its friendship-selection policy, or editing the graph it generates. Across Llama and Gemma, PPO brought mean political and race/ethnicity homophily closer to their empirical targets during generation. Applying TADP to the PPO outputs then brought all four mean demographic ratios within 0.03 of their targets while preserving each persona's degree. Starting from PPO rather than Base outputs required 32.7\% fewer swaps for Llama and 17.5\% fewer for Gemma. Although both starting points yielded similar final demographic mixing, the share of original friendships retained was higher with PPO by 11.46 percentage points for Llama and 5.82 percentage points for Gemma.

\clearpage
\bibliographystyle{IEEEtran}
\bibliography{paper_references}
\end{document}